\documentclass[dvips]{article}
\usepackage{amsmath,amssymb}
\usepackage{latexsym}
\usepackage{epsfig}
\usepackage{icrctc07}

\title{Effect of muon-nuclear inelastic scattering on high-energy atmospheric muon \\
 spectrum at large depth underwater}
 
\shorttitle{Effect of muon-nuclear inelastic scattering}
\authors{S.~I.~Sinegovsky$^{1}$, A.~Misaki$^{2}$, K.~S.~Lokhtin$^{1}$ and N.~Takahashi$^{3}$.}
     
\shortauthors{S.I. Sinegovsky et al.}     
\afiliations{$^1$Irkutsk State University, Irkutsk, Russia \\ 
             $^2$Advanced Research Institute for Science and Technology,
Waseda University, Tokyo, Japan \\
 $^3$ Department of Advanced Physics, Hirosaki University, Hirosaki, Japan}
\email{misakiakeo@air.ocn.ne.jp}

\abstract
{
 The energy spectra of hadron cascade showers produced by the cosmic ray muons travelling
through water as well as the muon energy spectra underwater at the depth up to $4$ km are
calculated with two models of muon inelastic scattering on nuclei, the recent hybrid
model (two-component, $2C$) and the well-known generalized
vector-meson-dominance model for the comparison. The $2C$ model involves
photonuclear interactions at low and moderate virtualities as well as the hard
scattering including the weak neutral current processes. For the muon scattering off
nuclei substantial nuclear effects, shadowing, nuclear binding and  Fermi motion of nucleons are taken into account. It is shown that deep underwater muon energy spectrum calculated
with the $2C$ model are noticeably distorted at energies above $100$ TeV as compared to
that obtained with the GVMD model.}

\begin{document}
\maketitle

\section{Introduction}

The muon inelastic scattering on nuclei contributes noticeably to the total energy loss of cosmic rays muons. The influence of this interaction on the shape of ultra-high energy muon spectra at the great depth of a rock/water is still unknown in detail. Of interest is also to estimate the number of cascade showers produced by very high-energy muons through inelastic interactions with nuclei, and to study the influence of this process on the energy spectra  of cosmic-ray muons in water at the depths of the underwater/ice neutrino telescopes -- NT200+, AMANDA, IceCube, ANTARES, NESTOR, NEMO and KM3NeT.  

In this work, we calculate energy spectra of hadron cascade showers produced by cosmic-ray (atmospheric) muons in water due to inelastic scattering on nuclei, as well as the integral energy spectra of atmospheric muons in water at depths up to $4$ km. Calculations are performed with two models: the hybrid model of inelastic scattering of leptons on nuclei~\cite{KLS2004, KLS2005} and, for a comparison, the known generalized vector-meson-dominance (GVMD) model of photonuclear muon interactions by Bezrukov and Bugaev~\cite{BB-81}.

\section{Muon-nucleus inelastic scattering} 

The  hybrid two-component ($2C$) model~\cite{KLS2004, KLS2005} for inelastic scattering of high-energy charged leptons on nuclei involves photonuclear interactions at low and moderate  $Q^2$ as well as the deep inelastic scattering processes at high $Q^2$ values.
For the virtuality $0<Q^2\leq 5$ GeV$^2$ the Regge based parametrization~\cite{CKMT} for the electromagnetic structure function $F_2^{\gamma}$ is applied and the muon-nucleon inelastic scattering cross section at $Q^2\leq 5$ GeV$^2$ is computed with the formula
\begin{eqnarray}
\label{CS_CKMT} \nonumber 
 \frac{d^2\sigma}{dQ^2dy}=\frac{4\pi\alpha^2}{yQ^4}F_2^{\gamma}(x,Q^2)\left[1-y-\frac{Q^2}{4E^2}
	 + \right. \\ 
 \left.\frac{y^2}{2(1+R)}\left(1-\frac{2m_\mu^2}{Q^2}\right)
\left(1+\frac{Q^2}{E^2y^2}\right)\right],
\end{eqnarray}
where the ratio $R=\sigma_L/\sigma_T$ is taken into account according to Ref.~\cite{Abe},
 $y=\nu/E$ is the fraction of muon energy transfered to the hadron system, $Q^2$ and $x=Q^2/(2MEy)$ are the Bjorken variables. 
 \begin{table*}[ht!]%%%%%%%%%%%%%%%%%%%%%%%%%%%%%%%%%%%%%%%%%%%%%%%%%%%%%%%%%%%%%%
\caption{Comparison of calculations of the muon energy loss due to inelastic scattering in standard rock. %calculated with different models. 
\label{tab_comp}}
%of muon-nucleus inelasic scattering.
\begin{center}
\begin{tabular}{c|ccccc} \hline\hline
 $E$, &\multicolumn{5}{c}{${b_n(E),\ 10^{-6}}$ cm$^2\cdot$ g$^{-1}$}\\
 GeV  &
\qquad \cite{KLS2004,KLS2005}&
\qquad\cite{Dutta}  &
\qquad\cite{BM}     &
\qquad\cite{BSh}    &
\qquad\cite{PT}      \\ 
  & \qquad ($2C$)  &  &  &    \\ 
\hline
%&\multicolumn{5}{c}{} \\
$10^5$&\qquad 0.62 \ &\qquad 0.60&\qquad 0.68&\qquad 0.70 &\qquad 0.70 \\
$10^6$&\qquad 0.82 \ &\qquad 0.80&\qquad 0.88&\qquad 1.08 &\qquad 1.00 \\
$10^8$&\qquad 1.53 \ &\qquad 1.60&\qquad $-$ &\qquad 2.25 &\qquad 2.50 \\
$10^9$&\qquad 2.16 \ &\qquad 2.18&\qquad $-$ &\qquad 3.10 &\qquad 4.00 \\
%&\multicolumn{5}{c}{} \\
\hline\hline
\end{tabular}
\end{center}
\end{table*}%%%%%%%%%%%%%%%%%%%%%%%%%%%%%%%%%%%%%%%%%%%%%%%%%%%%%%%%%%%%%%%%%%% 

In the  range $Q^2> 5$ GeV$^2$ the cross section of $\mu N$-scattering can be written in the form\begin{eqnarray}
\label{dS_dydQ2}\nonumber
\frac{d^2\sigma}{dQ^2dy}=\frac{4\pi\alpha^2}{yQ^4}
\left[F_2^{NC}\left(1-y-\frac{Q^2}{4E^2}+\right.\right. \\ 
\left.\frac{y^2}{2}-\left.\frac{y^2m_\mu^2}{Q^2}\right)
\pm\left(\frac{y^2}{2}-y\right)xF_3^{NC}\right],
\end{eqnarray}
where we put  $R=Q^2/\nu^2$, that is equivalent to the Callan-Gross
relation, $F_2=2xF_1$. Signs ``$\pm$'' stand for $\mu^{\pm}$.
In Eq. (\ref{dS_dydQ2}) used notations are:
\begin{align} 
\label{F2_NC}  \nonumber
&F_2^{NC}=F_2^{\gamma}-g_V^\mu \eta_{\gamma Z}F_2^{\gamma
  Z}+({g_V^\mu }^2+{g_A^\mu }^2)\eta_{\gamma Z}^2 F_2^Z,&  \\ %\nonumber
&F_3^{NC}=-g_A^\mu \eta_{\gamma Z}F_3^{\gamma Z}+2g_V^\mu g_A^\mu \eta_{\gamma Z}^2 F_3^Z;& \\
% \end{align}
%\begin{align} 
&\eta_{\gamma Z}=\frac{G_FM_Z^2}{2\sqrt{2}\pi\alpha}
 \frac{Q^2}{M_Z^2+Q^2},& \\ 
& g_V^\mu =-\frac{1}{2}+2\sin^2\theta_W, \  g_A^\mu =-\frac{1}{2}.&
 \end{align}
The structure functions (SFs) $F_2^Z$, $F_3^Z$  represent the weak neutral current (NC) contribution, $F_2^{\gamma Z}$, $F_3^{\gamma Z}$ are taking  into account the electromagnetic and weak current interference. The nucleon SFs, $F_2^\gamma$, $F_2^{\gamma Z}$, $F_2^{Z}$, $F_3^{\gamma Z}$, $F_3^Z$,  are defined in the quark-parton picture through the parton distributions (see \cite{Anselm94, PDG04}).
For the range of $Q^2>6$ GeV$^2$ the electroweak nucleon SFs
are computed with the CTEQ6~\cite {CTEQ6} and MRST~\cite{MRST-02} sets of the parton
distributions. Linear fits for the nucleon SFs are used in the range $5< Q^2<6$ GeV$^2$. 

Nuclear modifications of the nucleon SFs due to coherent and incoherent effects~\cite{Arneodo94, Piller00} are taken into account according to Ref.~\cite{smirnov95} (see also~\cite{BM}) by the factor $r_A(x,Q^2)=F_2^A/F_2^N$ that is the ratio of the structure function per nucleon of a nucleus with mass number $A$ and the isospin averaged nucleon structure function $F_2^N$. Thus,
\begin{equation}\label{dsA_dy}
\frac{d\sigma^{\mu A}(E,y)}{dy}= 
A\int\limits^{Q^2_{\rm max}}_{Q^2_{\rm min}}dQ^2\,r_A\frac{d^2\sigma}{dQ^2dy}.
\end{equation} 
The contribution of muon-nuclear interactions to the continous energy loss is given by integral
\begin{equation}\label{b_N}
b_n(E)\equiv-\frac{1}{E}\frac{dE}{dh}=N_0
\int\limits^{y_{\rm max}}_{y_{\rm min}}dy\,y\frac{d\sigma^{\mu A}}{dy},
\end{equation}
where $N_0= N_A/A$ is the number of nuclei per gramme of matter.

Table~\ref{tab_comp} presents the muon energy loss $b_n$ due to muon-nucleus interactions in rock, computed for various models:  values of $b_n$  obtained for the $2C$ model are given in the second column, the rest are results predicted in Refs.~\cite{BM, Dutta, BSh, PT}. One may see  that the muon energy loss $b_n(E)$ at $E>10^6$ GeV obtained in Refs.~\cite{KLS2004, KLS2005, Dutta} differ apparently from the predictions~\cite{BSh, PT}.

The energy dependence of the inelastic scattering energy loss for muons traveling
through water may be parametrized in  the energy range $10^2-10^9$ GeV as
 %($\eta =\lg({E}/1 \,{\rm GeV})$ :
\begin{eqnarray} \label{fit_b}
b_n (E) =c_0+c_1\eta+c_2\eta^2+c_3\eta^3+c_4\eta^4,
 \end{eqnarray}
where $\eta =\log_{10}({E}/1 \,{\rm GeV})$ and coeffecients $c_i$ (in units of $10^{-6}$ \ {\rm cm$^2$g$^{-1}$})  are  
\begin{align}
 c_0&= 1.06416, \quad c_1=-0.64629,\, c_2=0.20394, \nonumber \\
 c_3&=-0.02465, \, c_4=0.00130. 
\end{align}

\begin{figure}[ht!]
\begin{center}
\includegraphics[width=0.48\textwidth]{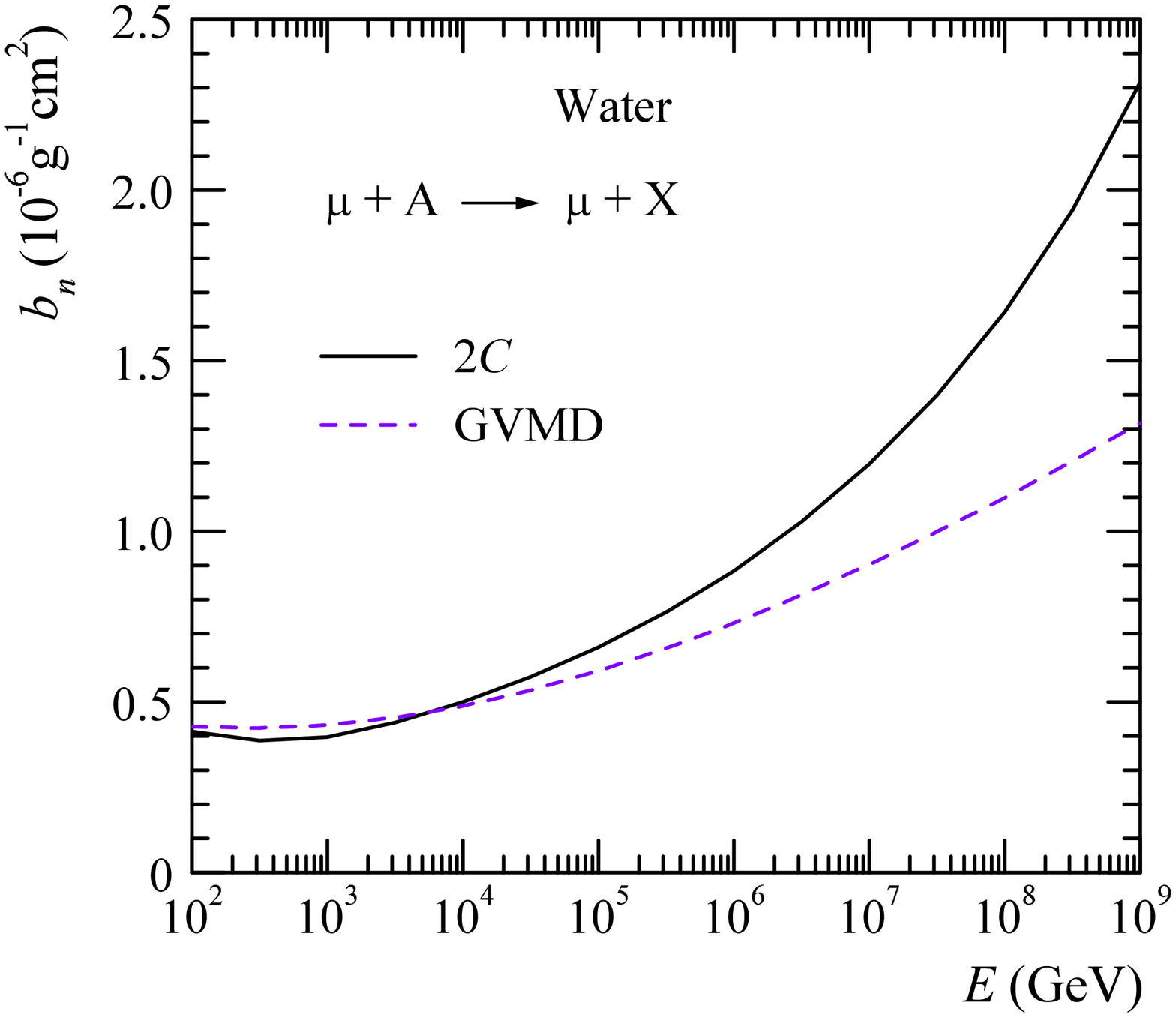}
\end{center}
\vskip -6 mm \caption{The muon energy loss due to muon-nucleus interactions in water.}
\label{bmutau}
% calculated with the $2C$ model (solid line) and the GVMD one (dashed)loss due to
%$\ell^\pm$-nucleus inelastic scatering in water.} \label{delta-bw}
\end{figure}

Figure~\ref{bmutau} shows the energy loss $b_n(E)$ for the inelastic scattering of muons in water calculated with the  $2C$ model (solid curve) and the GVMD one (dashed).

\section{Muon induced hadron showers in water}

\begin{figure}[t]
\begin{center}
\includegraphics[width=0.48\textwidth]{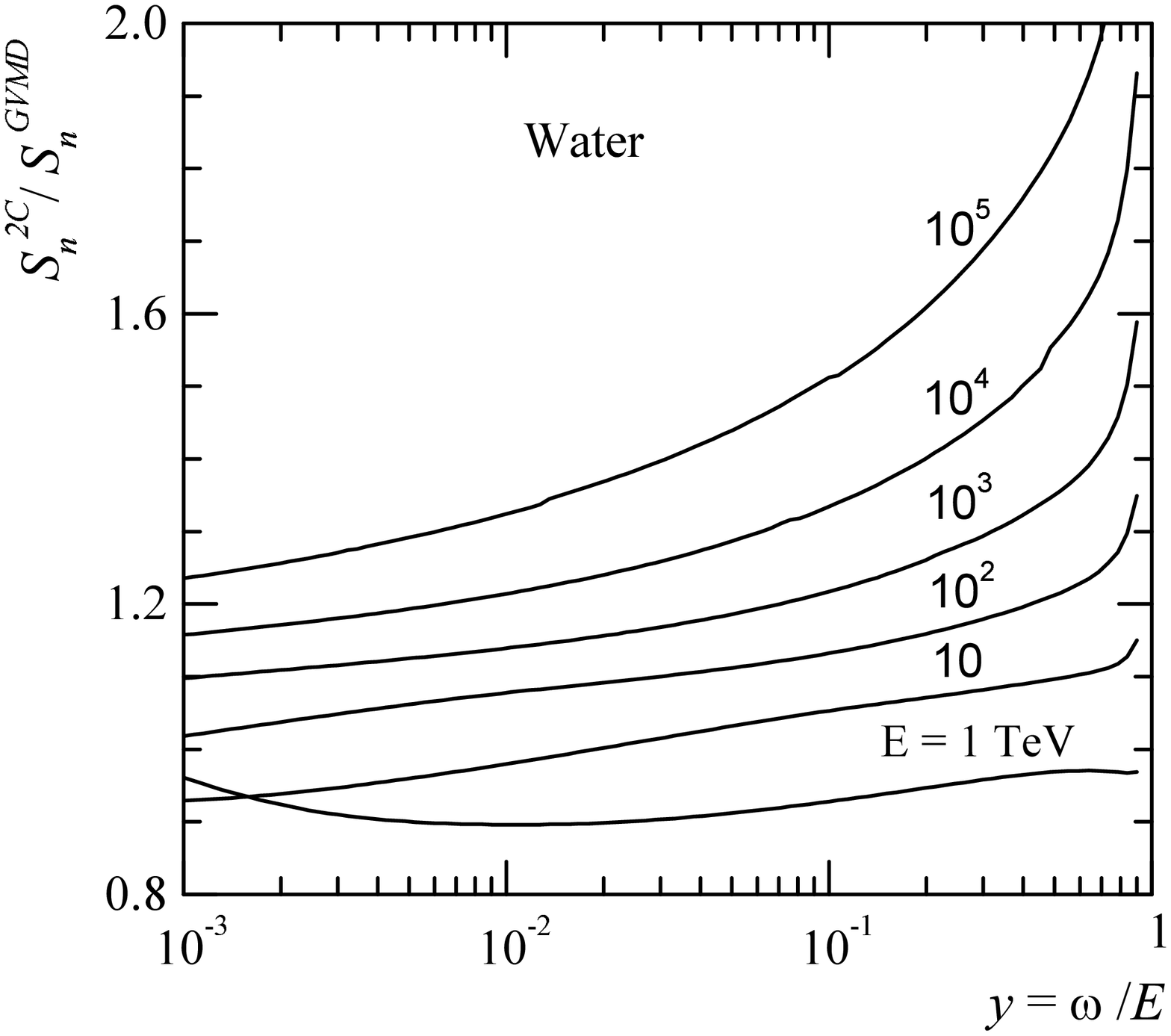}
\end{center}
\vskip -6 mm \caption{The ratio of the energy spectra of hadron showers computed with the $2C$ model to those obtained with use of the GVMD one.} \label{shower}
\end{figure}

The number of hadron showers with energies above $\omega $  per cm$^2$ per second per steradian, generated in water column $\Delta h=h_2-h_1$ through muon-nucleus interactions at   energy above $E$, may be defined as
\begin{align}\nonumber
	S_n(\omega,E,\Delta h,\theta)=&N_0 \int\limits_{h_1}^{h_2}dh\int\limits^{\infty}_{E}d\varepsilon\,D_\mu(\varepsilon,h,\theta)\times& \\
 &\int\limits^{y_{\rm max}}_{\omega/\varepsilon}dy\,
 \frac{d\sigma^{\mu A}(\varepsilon,y)}{dy}.& 
\end{align}
Here $D_\mu(E,h,\theta)$ is the muon flux  (cm$^{-2}$s$^{-1}$sr$^{-1}$GeV$^{-1}$) at depth $h$,   $\theta$ is zenith angle, $y_{\rm max}= 1- m_\mu/E$.
Ratios of the shower spectra computed with the $2C$ model to those obtained with the GVMD one are shown in figure~\ref{shower}. At the muon energy $E = 10$ TeV numbers of the muon-induced hadron showers, calculated with two models, the $2C$ and GVMD, differ slightly (about $10$\%). 
However the discrepancy between the models grows with increasing energy:
 for $E = 100$ TeV, it is as great as $\sim 30$\,\%, and for $E = 10^5$ TeV, the result obtained for the $2C$ model in the range of the large energy transfer 
exceeds by a factor of $\sim 2.5$ the corresponding result for the model of photonuclear interaction~\cite{BB-81}. 
For energies $E > 10^3$ TeV, the number of showers calculated with the $2C$ model exceeds  that obtained with use of the GVMD model by $20$–-$50$\,\% even in the region of small energy loss ($y \sim 0.1$). For the catastrophic energy loss ($y > 0.5$), the
number of showers, obtained with use of the $2C$ model, exceeds that of the GVMD prediction by factor about $2$. 

\section{Atmospheric muon spectra underwater}

The solution of the muon transport equation~\cite{NSB94} allows to obtain the differential energy spectra $D_\mu(E,h,\theta)$ at large depth in homogeneous media and therefore to compute the  flux  of cosmic ray muons with energy above specified one: 
\begin{equation}
	N_\mu(E, h,\theta)=\int\limits_E^\infty d\varepsilon \, D_\mu(\varepsilon,h,\theta).
\end{equation}

\begin{figure}[t!] %%%%%%%% %\vskip -5mm \hskip -5mm
\vskip -2mm
\begin{center}
\includegraphics[width=0.49\textwidth]{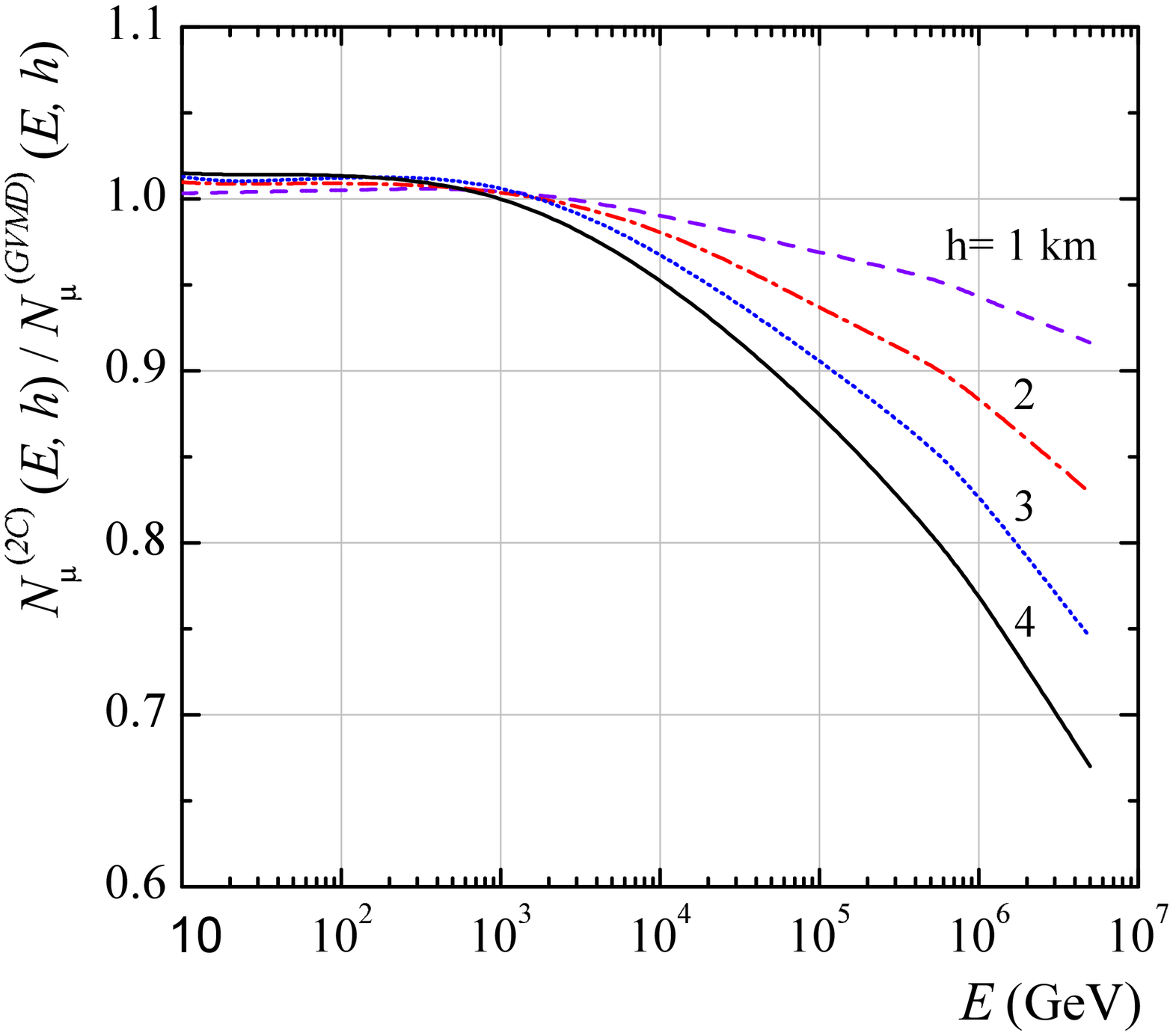}
\end{center}
 \vskip -5 mm %height=6.cm
 \caption{The ratio of muon fluxes underwater calculated with the $2C$ model to those obtained with the GVMD one.}
\label{flux_rat}
\end{figure}

Figure~\ref{flux_rat} shows the ratio $N^{2C}_\mu(E, h)/N^{GVMD}_\mu(E, h)$ of the near vertical muon flux underwater computed at depth $1$--$4$ km  for two models of muon-nuclear scattering, the 2C model~\cite{KLS2005} and GVMD one~\cite{BB-81}. The effect is rather noticeable at the $E>10$ TeV, especially for the depth $3$--$4$ km. For $h=4$ km this ratio decreases to $0.75$ at $E=10^3$  TeV. Thus a sizeable increase of the muon inelastic scattering cross section  may  result in an appreciable decrease of the deep underwater muon flux as compared to that obtained~\cite{prd58, prd63} for the GVMD model. It should be noted that this result refers only to the atmospheric conventional ($\pi, K$) muons. As concerns muons produced in charmed particle decays (prompt muons), which become presumably dominant at $E > 10^5$ GeV (see e. g.~\cite{prd58,prd63}), the role of the muon-nucleus inelastic scattering needs further study. 
 
\section{Conclusions}
 
Evidently the increase of the cross section of inelastic muon scattering in matter, while leading to diminished cosmic-ray muon flux deep underwater, results in growing efficiency of muon registration. This last factor is positive for neutrino astronomy since neutrino-induced muons may yield the signal from astrophysical high-energy muon neutrinos. 

\section{Acknowledgements}
S.~Sinegovsky acknowledges the support by Federal Program  "Leading Scientific Schools of Russian Federation",  grant NSh-5362.2006.2.


\begin{thebibliography}{99}

\bibitem{KLS2004}
K.~S.~Kuzmin, K.~S.~Lokhtin, S.~I.~Sinegovsky, Int. J. Mod. Phys. A \textbf{20}, 6956 (2005);  hep-ph/0412377.

\bibitem{KLS2005}
A.~A.~Kochanov,  K.~S.~Lokhtin and S.~I.~Sinegovsky,  in {\it Proc. of 29th ICRC, Pune, 2005},  Vol. 9, p. 69; hep-ph/0508306.

\bibitem{BB-81}
L.~B.~Bezrukov and E.~V.~Bugaev,  Sov. J. Nucl. Phys. \textbf{33}, 635 (1981).

\bibitem{CKMT}
A.~Capella \emph{et al.},  Phys. Lett.  B \textbf{337}, 358 (1994); 
%%CITATION = HEP-PH 9405338;%%
%\bibitem{KMP}
A.~B.~Kaidalov, C.~Merino and D.~Pertermann,  Eur. Phys. J. C \textbf{20}, 301 (2001).
%%CITATION = EPHJA,C20,301;%%

\bibitem{Abe}
 K.~Abe \emph{et al.}, Phys. Lett. B \textbf{452}, 194 (1999).%;  hep-ex/9808028.
% MEASUREMENTS OF R = SIGMA(L) / SIGMA(T) FOR 0.03<X<0.1 AND FIT TO WORLD DATA.
% By E143 Collaboration (K. Abe et al.). SLAC-PUB-7927, Aug 1998. 8pp.
% Published in Phys.Lett.B452:194-200,1999
% e-Print Archive: hep-ex/9808028

\bibitem{Anselm94}
 M.~Anselmino, A.~Efremov and E.~Leader,
 Phys. Rept. \textbf{261}, 1 (1995).
%  Erratum-ibid.\textbf{281}, 399 (1997).
 
\bibitem{PDG04}  
  S.~Eidelman \emph{et al.}, Phys. Lett. B \textbf{592}, 1 (2004).

\bibitem{CTEQ6} 
J.~Pumplin \emph{et al.},  JHEP \textbf{0207}, 012 (2002).%;  hep-ph/0201195.
%%CITATION = HEP-PH 0201195;%%

\bibitem{MRST-02}
 A.~D.~Martin \emph{et al.},   Eur. Phys. J. C \textbf{23}, 73 (2002).
%%CITATION = EPHJA,C23,73;%%
%Journal-ref: Eur.Phys.J. C23 (2002) 73-87

\bibitem{Arneodo94} M. Arneodo,  Phys. Rep. \textbf{240}, 301 (1994).
%%CITATION = PRPLC,240,301;%%

\bibitem{Piller00}
G.~Piller  and  W.~Weise, Phys. Rep. \textbf{330}, 1 (2000).
 
\bibitem{smirnov95}
 G.~I.~Smirnov, Phys. Lett. B \textbf{364}, 87 (1995); \\
 G.~I.~Smirnov, Eur. Phys. J. C \textbf{10}, 239 (1999).

\bibitem{BM} A.~V.~Butkevich and S.~P.~Mikheyev,  J. Exp. Theor. Phys. \textbf{95}, 11 (2002); hep-ph/0109060.
%%% A. V. Butkevich, S. P. Mikheyev, {\it J. Exp. Theor. Phys.} {\bf 95}, 11 (2002).
%%CITATION = HEP-PH 0109060;%%

\bibitem{Dutta}
 S.~I.~Dutta \emph{et al.},  Phys. Rev. D \textbf{63}, 094020 (2001).
%%CITATION = PHRVA,D63,094020;%%


\bibitem{BSh} E.~V.~Bugaev and Yu.~V.~Shlepin, Phys. Rev. D \textbf{67}, 034027 (2003); hep-ph/0203096 v5.
%%CITATION = PHRVA,D67,034027;%%

\bibitem{PT}
A.~A.~Petrukhin A.A. and D.~A.~Timashkov, Phys. Atom. Nuc.  \textbf{67},  2216 (2004); 
D.~A.~Timashkov  and A.~A.~Petrukhin,  in {\it Proc. 29 ICRC. Pune, 2005}, Vol. 9, p. 89.

\bibitem{NSB94}
V.~A.~Naumov, S.~I.~Sinegovsky and E.~V.~Bugaev, 
Phys. Atom. Nucl. \textbf{57},  412 (1994); hep-ph/9301263.
%Yad. Fiz., 57, 439–451 (1994)

\bibitem{prd58}
E.~V.~Bugaev \emph{et al.}, Phys. Rev. D \textbf{58}, 054001 (1998); hep-ph/9803488.

\bibitem{prd63}
T.~S.~Sinegovskaya and S.~I.~Sinegovsky, Phys. Rev. D \textbf{63}, 096004 (2001); hep-ph/0007234.


\end{thebibliography}
\end{document}